\documentclass[aip, jcp, amsmath,amssymb, reprint]{revtex4-1}
\usepackage{graphicx}
\usepackage[utf8]{inputenc}
\usepackage[T1]{fontenc}
\usepackage{mathptmx}
\usepackage{etoolbox}
\usepackage{dcolumn}
\usepackage{xcolor,hyperref}
\hypersetup{
   colorlinks,
   linkcolor={red!80!black}, 
   citecolor={blue!50!black},
   urlcolor={blue!80!black}
}

\makeatletter
\def\@email#1#2{%
 \endgroup
 \patchcmd{\titleblock@produce}
  {\frontmatter@RRAPformat}
  {\frontmatter@RRAPformat{\produce@RRAP{*#1\href{mailto:#2}{#2}}}\frontmatter@RRAPformat}
  {}{}
}%

\makeatother

\begin{document}

\title{Temperature Dependence of Thermodynamic, Dynamical and Dielectric Properties of Water Models}

\author{Tatiana I. Morozova}
\affiliation{Institut Laue-Langevin, 71 Avenue des Martyrs, 38042 Grenoble, France}
\email{morozova@ill.fr}
\author{Nicol\'as A. Garc\'ia}
\affiliation{Departamento de Física, IFISUR-UNS-CONICET, 8000 Bahía Blanca, Argentina}
\author{Jean-Louis Barrat}
\affiliation{Institut Laue-Langevin, 71 Avenue des Martyrs, 38042 Grenoble, France}
\affiliation{Univ. Grenoble Alpes, CNRS, LIPhy, 38000 Grenoble, France}


\begin{abstract}
We investigate the temperature dependence of thermodynamic (density, isobaric heat capacity), dynamical (self-diffusion coefficient, shear viscosity), and dielectric properties of several water models, the commonly employed TIP3P water model, well-established 4-point water model TIP4P-2005, and recently developed 4-point water model TIP4P-D. We focus on the temperature range of interest for the field of computational biophysics and soft matter (280-350 K). The 4-point water models lead to a spectacularly improved agreement with experimental data, strongly suggesting that the use of more modern parametrizations should be favored compared to the more traditional TIP3P for modeling temperature-dependent phenomena in biomolecular systems.
\end{abstract}

\maketitle 

Due to its unique properties such as high dielectric constant, hydrogen bonding structure, mobility, water has unique solvent properties. For instance, water plays a crucial role as a solvent in many biological processes. Among them are enzymatic catalysis, \cite{adkar2011role} self-organization from single- to double-stranded DNA,\cite{cui2006weakly} and the phase behavior of intrinsically disordered proteins (IDPs) \cite{shin2017liquid} - proteins that lack stable secondary and tertiary structures. To investigate such phenomena computationally, it is required to probe several temperatures. Thus, the water model employed should reproduce the properties of liquid water for a  range of physiological temperatures, and in fact, the influence of the model used for water molecules has been noticed in the literature (see for example  Ref. \citenum{emperador2021effect}). 
However, this is not generally related to the intrinsic properties of the water model.  In fact, these models are often parametrized for a given temperature, and it is not \textit{a priori} guaranteed that their performance at other temperatures will be satisfying. Here we investigate the temperature dependence of thermodynamic, dynamical, and dielectric properties of three water models, namely the TIP3P model \cite{jorgensen1983comparison} commonly used in the field of computational biophysics, TIP4P-2005 model \cite{abascal2005general} which reproduces well the behaviour of real water, and a recently developed TIP4P-D model \cite{piana2015water} with increased repulsive and dispersion interaction. The latter was developed to improve the description of structural properties of IDPs and adopted the geometry of TIP4P-2005 model. We focus on the liquid water state as of interest for the field of computational biophysics and soft matter. To our knowledge, this is the first time when liquid properties of the TIP4P-D water model are reported as a function of temperature.

\begin{figure*}[htp]
    \centering
    \includegraphics{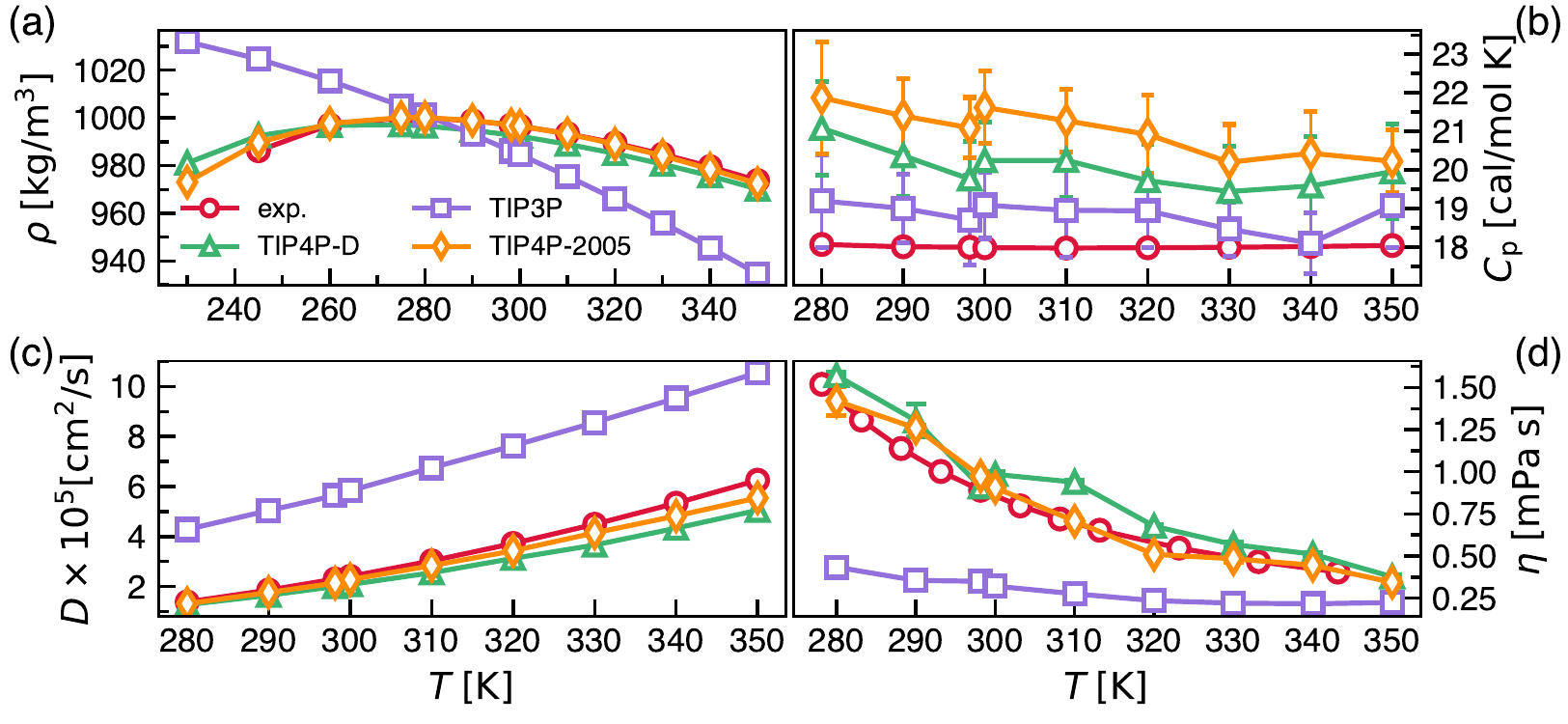}%
    \vspace{-0.5cm}
    \caption{Thermodynamic and dynamical properties of three water models TIP4P-D, TIP4P-2005, and TIP3P as a function of temperature: (a) density $\rho$, (b) isobaric heat capacity $C_{\rm p}$ (values do not include vibrational corrections), \cite{horn2004development} (c) self-diffusion coefficient $D$, and (d) shear viscosity $\eta$. The simulation results are compared with the corresponding experimental values. For the graphs where the error bars are not shown the standard error is smaller than the symbol size.}%
    \label{fig:all_prop}
\end{figure*}
All simulations have been conducted using the GROMACS molecular dynamics (MD) package (ver. 2020.1).\cite{abraham2015gromacs} Simulations consist of $N=4000$ water molecules in a cubic box with box length $L\simeq 5$ nm. We consider a temperature range from $T=280$ K to $350$ K with a step of $10$ K. We also include a value for room temperature, i.e. $T=298.15$ K. At each temperature, we perform a two-step system equilibration by first conducting a short $NVT$  (0.1 ns) and  $NpT$ (1 ns) warmup simulations at pressure equaled to 1 atm. We then conduct three independent production runs of 30 ns in the $NpT$ ensemble for computing the average density $\rho$ and the isobaric heat capacity $C_{\rm p}$. We compute the isobaric heat capacity using the enthalpy fluctuation as $C_{\rm p} = 1/N (\langle H^2 \rangle - \langle H \rangle^2 )/ k_{\rm B} \langle T \rangle^2$, where $H$ is the enthalpy of the system, $k_{\rm B}$ is the Boltzmann constant. To broaden the scope of our study, we extend it to the supercooled water regime by also investigating the density variation in the temperature range 230 K - 275 K. At these temperatures, the simulation length is increased to 100 ns.

We calculate the self-diffusion coefficient $D$ and the static dielectric constant $\varepsilon(0)$ in the $NVT$ ensemble at the average density $ \langle \rho(T) \rangle$ by conducting three independent production runs of 30 ns. We estimate the self-diffusion coefficient from the mean-squared displacement using the Einstein relation as $D=\lim_{t\to\infty} 1/6 \left (d \langle | \mathbf{r}(t) - \mathbf{r(0)} | \rangle ^2/dt \right)$. Next, we compute the shear-viscosity via the Green-Kubo relation $\eta=V/k_{\rm B}T \int_{0}^{\infty} dt \langle p_{\alpha \beta} (t) p_{\alpha \beta}(0) \rangle$, where $V$ is the system volume, $p_{\alpha \beta}$ are the Cartesian components of the stress tensor. We improve the sampling by using all five independent components of the traceless stress tensor: $p_{xy}$, $p_{yz}$, $p_{zx}$, $1/2(p_{xx} - p_{yy})$, and $1/2(p_{yy} - p_{zz})$. \cite{alfe1998first} It was previously shown that the auto-correlation function (ACF) of the shear stress decays on the time-scale of a few picoseconds in liquid water. \cite{tazi2012diffusion} Thus, we perform at least 18 short simulations in the $NVT$ ensemble, each of 20 ps, where the pressure tensor of the system is saved every $2$ femtoseconds. We compute an ACF averaged over all realizations. Error bars correspond to the standard deviation from fitting $\eta(t)$ with $t>2$ ps to a constant.

In simulations conducted in the $NpT$ ensemble, the pressure is kept constant at $p$=1 atm using the Parrinello-Rahman barostat \cite{parrinello1980crystal} with a coupling constant of 2 ps, and the temperature is kept constant by a velocity rescaling thermostat \cite{bussi2007canonical} with a coupling constant of 1 ps. For simulations in the $NVT$ ensemble, the temperature is held by the Nos\'e-Hoover thermostat \cite{nose1984molecular, hoover1985canonical} with the coupling constant of 0.5 ps. The electrostatic interactions are simulated using the particle mesh Ewald (PME) algorithm. \cite{essmann1995smooth} The cutoff for the electrostatic and van der Waals interactions is set to 1.2 nm. Van der Waals interactions outside the cutoff distance are approximated via a continuum model (DispCorr=EnerPres). Bonds are constrained using the LINCS algorithm. \cite{hess1997lincs} The equations of motion are integrated using the leap-frog integrator with a time step of 2 fs. Note that we do not follow the MD setting for which the water models were initially parametrized. Instead, we employ those  (e.g. PME) which are typically used in studies on polymer/protein solutions. We analyze the trajectories using in-house scripts as well as GROMACS analysis tools such as $gmx \; energy$ ($\rho$, $H$, $p_{\alpha \beta}$), $gmx \; msd$ ($D$), and a python library MDTraj ($g(r)$,$\varepsilon(0)$). \cite{mcgibbon2015mdtraj}

\begin{figure}
    \includegraphics{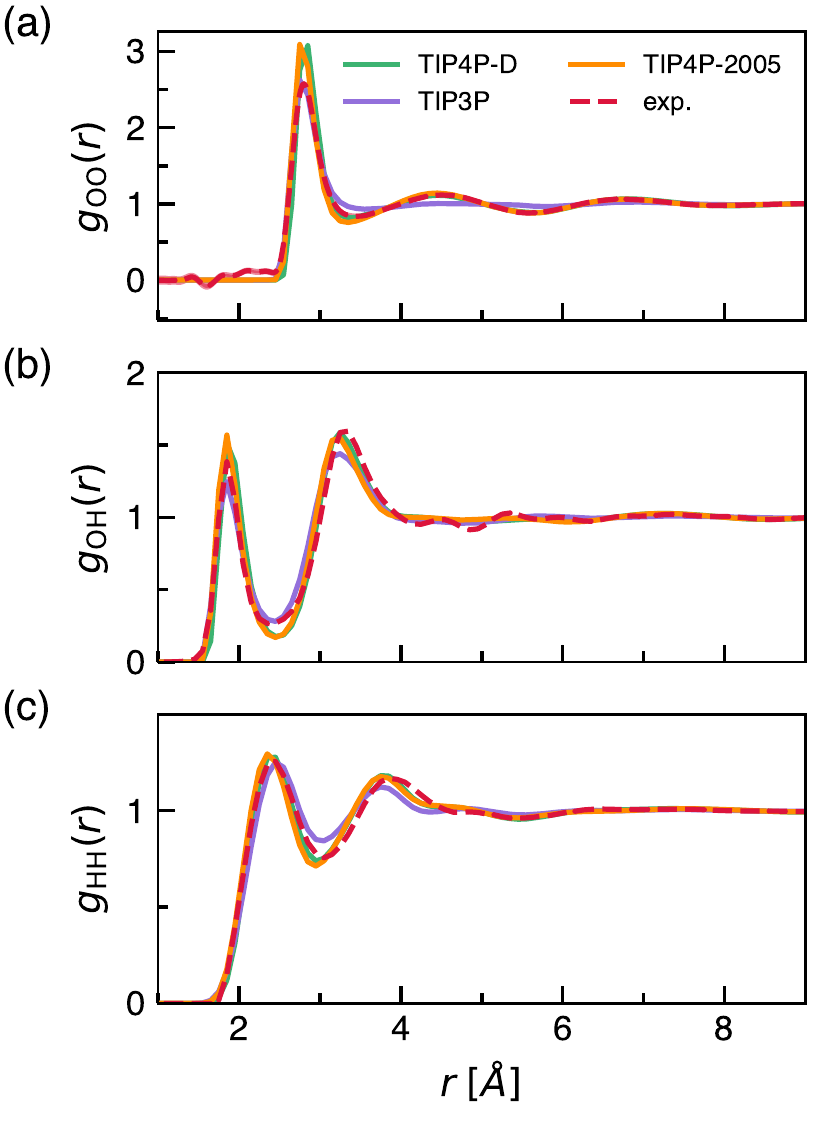}
     \vspace{-0.5cm}
    \caption{Intermolecular radial distribution functions of three water models TIP4P-D, TIP4P-2005, and TIP3P at 298.15 K and 1 atm: (a) the oxygen-oxygen $g_{\rm {OO}}(r)$, (b) the oxygen-hydrogen $g_{\rm {OH}}(r)$, and (c) the hydrogen-hydrogen $g_{\rm {HH}}(r)$ radial distribution functions. Experimental results of liquid water are presented as well. } %
    \label{fig:gofr}
\end{figure}

\begin{figure}
    \includegraphics{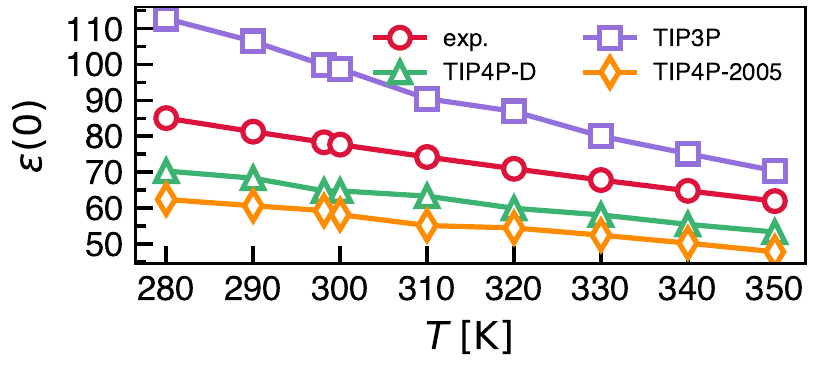}%
    \vspace{-0.5cm}
    \caption{Static dielectric constant $\varepsilon (0)$ of three water models TIP4P-D, TIP4P-2005, and TIP3P as a function of temperature. Experimental values of $\varepsilon (0)$ for corresponding temperatures are also shown.}%
    \label{fig:dielec_const}
\end{figure}

In Fig.~\ref{fig:all_prop} we plot the temperature dependence of thermodynamic and dynamical properties of liquid water. In Fig.~\ref{fig:all_prop}(a), we compare the experimental (exp) density \cite{kell1975density,wagner2002iapws} as a function of the temperature with the water models. We do not include the value for the density at $T=230$ K in the experimental curve since at such temperature $\rho$ of the liquid water can not be reliably determined.\cite{kell1975density} We find an excellent agreement between experiments and TIP4P-2005 water model which was parametrized to reproduce densities of liquid water at ambient conditions. TIP4P-D water also captures well $\rho(T)$-dependence with a maximum of less than 0.6\% of error ($T=245$ K) highlighting a good performance of this water model for a wide temperature range. Results for TIP3P are only acceptable  for a small temperature window around the room temperature, while strongly deviating from the experimental values at higher and lower $T$.

In Fig.~\ref{fig:all_prop}(b), we plot the experimental and computed values of $C_{\rm p} (T)$. For all systems, $C_{\rm p}$ is roughly constant within the error bars for the $T$-range considered. The computed values of $C_{\rm p}$ exceed the experimental ones \cite{wagner2002iapws} at all temperatures, with the TIP3P model showing a slightly better agreement.

In Fig.\ref{fig:all_prop}(c), we report the self-diffusion coefficient as a function of temperature. Experimental values are taken from Ref.~\citenum{holz2000temperature}. The dynamics of water is correctly modeled by TIP4P-D and TIP4P-2005 models, while TIP3P strongly overestimates water diffusion. The latter issue was previously reported in several studies. \cite{vega2009ice, wang2014building}

As for the viscosity, the TIP4P-D and TIP4P-2005 models are in good agreement with the experiments, \cite{collings1983high} where the computed values slightly exceed the experimental ones as shown in Fig.~\ref{fig:all_prop}(d). The viscosity of the TIP3P model shows only a small $T$-dependence over the range investigated, where $\eta$ values are several times smaller than observed in experiments, consistent with faster self-diffusion.

Additionally, we investigate the structure of the liquid water at $T$=298.15 K and $p$=1 atm by computing the intermolecular radial distribution functions (RDFs). In Fig. \ref{fig:gofr} (a) we plot the oxygen-oxygen RDF $g_{\rm OO}(r)$ for all models investigated as well as the experimental results.\cite{skinner2013benchmark} TIP4P-D and TIP4P-2005 water models accurately reproduce the position of the first peak, while its height is overestimated by these models resulting in a over structured liquid. The position and the height of the other peaks are well captured by 4-point models. In contrast, TIP3P model only captures the height and the position of the first peak and fails to describe the structure of the liquid water beyond it. All water models yield overall good agreement with the experimental curves \cite{soper1986new} for oxygen-hydrogen $g_{\rm OH}(r)$ and hydrogen-hydrogen $g_{\rm HH}(r)$ RDFs  shown in Fig. \ref{fig:gofr} (b) and (c), respectively. However, 4-point water models better capture the height of the RDF peaks compared to TIP3P model.

Since interaction between a solute and solvent molecules in biological systems often includes electrostatic interactions, we also investigate the dielectric properties of the models. We compute the static dielectric constant as $\varepsilon(0)= 1 + 4\pi/3V k_{\rm B}T (\langle M ^2 \rangle - \langle M \rangle^2)$, where $M$ is the system's total dipole moment. In Fig.~\ref{fig:dielec_const},  we compare the computational and experimental values \cite{malmberg1956dielectric} for $\varepsilon (0)$. All models result in a deviation from  the experimental results, where values for the TIP3P model are too high, while the ones of 4-point models are too low. However, the values of $\varepsilon(0)$ for TIP4P-D model are closer to the experimental ones compared to TIP4P-2005 model. Additionally, the slope $d\varepsilon(0)/dT$ for the 4-point water models is much closer to the experimental value, which means that the changes in the electrostatic interaction with $T$ will be better reproduced for these models.  

We conclude that the commonly used TIP3P water model rather poorly reproduces thermodynamic, dynamical, structural, and dielectric properties of liquid water. Instead, TIP4P-D and TIP4P-2005 perform well in reproducing thermodynamic as well as dynamical properties. The latter are known to be vital for capturing correctly, in addition to phase behavior, the internal dynamics of macromolecules such as backbone and side-chain motions. 
In spite of their comparable performance in terms of bulk water properties, it was noted in previous work \cite{piana2015water} that TIP4P-D performs better than TIP4P-2005 for reproducing structural properties of disordered proteins at room temperature. It seems likely that the slight improvement in the dielectric constant for the TIP4P-D model also contributes to its better performance in modeling IDPs, which should therefore be maintained over the whole temperature range.

\vspace{-0.6cm}
\section*{Author Declarations}
\vspace{-0.6cm}
The authors have no conflicts to disclose.
\vspace{-0.6cm}
\section*{Data availability}
\vspace{-0.6cm}
The data that support the findings of this study are available from the corresponding author upon reasonable request.
\vspace{-0.6cm}
\section*{Credit line}
\vspace{-0.6cm}
This article may be downloaded for personal use only. Any other use requires prior permission of the author and AIP Publishing. This article appeared in \textit{J. Chem. Phys.} 156, 126101 (2022) and may be found at https://doi.org/10.1063/5.0079003. 
\vspace{-0.6cm}
\begin{acknowledgments} 
\vspace{-0.6cm}
This work was performed using HPC resources (GPU-accelerated partitions of the Jean Zay supercomputer) from GENCI–IDRIS (Grant 2021 - A0100712464).
\end{acknowledgments}
\vspace{-0.6cm}
\section*{References}
\vspace{-0.6cm}
\bibliography{references}

\end{document}